\documentclass[12pt,a4paper]{article}
\pdfoutput=1
\usepackage{amsmath}
\usepackage{color}
\usepackage{amsfonts}
\usepackage{amssymb}
\usepackage{graphicx}
\usepackage{geometry}
\usepackage{amssymb,epsfig,subfigure}
\usepackage{amssymb}
\usepackage{hyperref}
\usepackage{comment}
\usepackage[font=footnotesize]{caption}
\usepackage[T1]{fontenc}
\usepackage[utf8]{inputenc}
\usepackage{latexsym}
\usepackage[english]{babel}
\usepackage{cite}

\makeatletter
\renewcommand\section{\@startsection {section}{1}{\z@}%
                                 {-3.5ex \@plus -1ex \@minus -.2ex}
                                   {2.3ex \@plus.2ex}%
                                   {\normalfont\large\bfseries}}
\renewcommand\subsection{\@startsection{subsection}{2}{\z@}%
                                   {-3.25ex\@plus -1ex \@minus -.2ex}%
                                     {1.5ex \@plus .2ex}%
                                     {\normalfont\bfseries}}
\renewcommand\subsubsection{\@startsection{subsubsection}{3}{\z@}%
                                   {-3.25ex\@plus -1ex \@minus -.2ex}%
                                     {1.5ex \@plus .2ex}%
                                     {\normalfont\itshape}}
\makeatother

\def\pplogo{\vbox{\kern-\headheight\kern -29pt
\halign{##&##\hfil\cr&{\ppnumber}\cr\rule{0pt}{2.5ex}&\ppdate\cr}}}
\makeatletter
\def\ps@firstpage{\ps@empty \def\@oddhead{\hss\pplogo}%
  \let\@evenhead\@oddhead 
}
\def\maketitle{\par
 \begingroup
 \def\thefootnote{\fnsymbol{footnote}}
 \def\@makefnmark{\hbox{$^{\@thefnmark}$\hss}}
 \if@twocolumn
 \twocolumn[\@maketitle]
 \else \newpage
 \global\@topnum\z@ \@maketitle \fi\thispagestyle{firstpage}\@thanks
 \endgroup
 \setcounter{footnote}{0}
 \let\maketitle\relax
 \let\@maketitle\relax
 \gdef\@thanks{}\gdef\@author{}\gdef\@title{}\let\thanks\relax}
\makeatother

\numberwithin{equation}{section}

\newcommand\eea{\end{eqnarray}}
\newcommand\bea{\begin{eqnarray}}

\def\beq{\begin{equation}}
\def\eeq{\end{equation}}

\newcommand{\be}{\begin{equation}}
\newcommand{\ee}{\end{equation}}
\newcommand{\ba}{\begin{align}}
\newcommand{\ea}{\end{align}}
\newcommand{\bg}{\begin{gather}}
\newcommand{\eg}{\end{gather}}
\newcommand{\bseq}{\begin{subequations}}
\newcommand{\eseq}{\end{subequations}}

\usepackage{latexsym}
\usepackage{physics}
\usepackage{mathrsfs}
\usepackage{amsthm}
\usepackage{keyval}
\usepackage{ifthen}
\usepackage{amsbsy}      

\newcommand{\lag}{\mathcal{L}}
\newcommand{\ham}{\mathcal{H}}

\newcommand{\rrealt}{\mathbb{R}^3}
\newcommand{\vv}{{V}}

\newcommand{\traza}[1]{Tr\left({#1}\right)}

\newcommand{\as}{Y_{lm}}
\newcommand{\av}[1]{\overline{Y}_{lm}^{#1}}
\newcommand{\at}[1]{T_{lm}^{#1}}
\newcommand{\h}[1]{h_{l}^{#1}}
\newcommand{\ho}[1]{h_{0}^{#1}}
\newcommand{\hu}[1]{h_{1}^{#1}}
\newcommand{\drh}[1]{\partial_r h_{l}^{#1}}
\newcommand{\drho}[1]{\partial_r h_{0}^{#1}}
\newcommand{\drhu}[1]{\partial_r h_{1}^{#1}}

\newcommand{\pl}[1]{P_l^{#1}(\cos{\theta})}

\numberwithin{equation}{section}

\begin{document}
\setcounter{page}0
\def\ppnumber{\vbox{\baselineskip14pt
}}
\def\ppdate{
} \date{}

\author{Valentin Benedetti\footnote{e-mail: valentin.benedetti@gmail.com}, Horacio Casini\footnote{e-mail: casini@cab.cnea.gov.ar}\\
[7mm] \\
{\normalsize \it Centro At\'omico Bariloche and CONICET}\\
{\normalsize \it S.C. de Bariloche, R\'io Negro, R8402AGP, Argentina}
}

\bigskip
\title{\bf  Entanglement entropy of linearized \\ gravitons in a sphere
\vskip 0.5cm}
\maketitle

\begin{abstract}
We compute the entanglement entropy of a massless spin $2$ field  in a sphere in flat Minkowski space. We describe the theory with a linearized metric perturbation field $h_{\mu\nu}$ and decompose it in tensor spherical harmonics. We fix the gauge such that a) the two dynamical modes for each angular momentum decouple and have the dynamics of scalar spherical modes, and b) the gauge-fixed field degrees of freedom inside the sphere represent gauge invariant operators of the theory localized in the same region. In this way the entanglement entropy turns out to be equivalent to the one of a pair of free massless scalars where the contributions of the $l=0$ and $l=1$ modes have been subtracted. The result for the coefficient of the universal logarithmic term is $-61/45$ 
 and coincides with the one computed using the mutual information.    
\end{abstract}
\bigskip

\newpage

\tableofcontents

\vskip 1cm

\section{Introduction}

The entanglement entropy (EE) of vacuum fluctuations across a boundary in space    
has shown to be an interesting theoretical quantity in quantum field theory (QFT). The study of EE was  originally motivated by the quest to understand black hole entropy and entropy in gravity, but it turned out to have a more clear and natural formulation in QFT. Entropy in quantum mechanics is by definition a quantity associated to a state in an algebra of operators, and  ordinary QFT naturally comes with a built in correspondence of algebras with regions of the space.   

The situation in gravity is less clear precisely because it is not completely understood how ``regions'' in quantum gravity might be defined in terms of the operator content of the theory (see for example \cite{Camps:2018wjf,Donnelly:2017jcd}). Holographic theories give a simple, but perhaps only partial, answer, to this question. By restricting the region to a boundary region, the associated algebra is given by the one of the dual QFT in the boundary. Holographic EE \cite{Ryu:2006bv,Faulkner:2013ana} has shown there is a correspondence, at least at the semi-classical level, of this QFT entropy to an entropy in a gravity theory in the so called entanglement wedge \cite{ese1,ese2,ese3}.  

In the study of EE it is important to establish a correspondence of the different terms on the entropy with known physical quantities in the model. One such signature that allows us to distinguish models form their EE is given by the coefficient of the logarithmic term. In this sense there are in the literature several calculations of logarithmic corrections to the black hole entropy formula due to the EE of quantum fields in the semiclassical background, including gravitons (for a review see \cite{Solodukhin:2011gn,Sen:2012dw}). The graviton contribution may actually be of relevance to distinguish the gravity theory \cite{Sen:2012dw}.  
Nevertheless, logarithmic terms are subtle too. An example of the problems involved is the case of a Maxwell field. The logarithmic term for a free Maxwell field does not coincide with the expected trace anomaly \cite{Casini2016,dowkeresferagauge,huang}. However, the presence of electric or magnetic charges can change this result, no matter the mass of the charged particles \cite{ss1,ss2}. Without changing the theory to include charges this issue  has been discussed in the literature in an effective manner using the constructions of edge modes or extended Hilbert space (see for example \cite{don1,don2}).  
 
In this paper we compute the EE of free gravitons in flat space by treating the theory as a quantum field theory of helicity $2$ particles. In this sense we do not have to deal with the localization problems of a full quantum gravity theory. We show there are no conceptual issues for these free fields per se as QFT. As in the case of the Maxwell field, it is important in computing the EE to understand correctly what is the entropy one is computing, that is, what is the algebra and the state, as well as the meaning of the result in terms of the continuum theory. A natural way to do this is by interpreting the universal coefficients in terms of the mutual information. This is  transparent in the real time formulation that we use in this paper where we have the quantum degrees of freedom always in sight. Computations using the replica trick may actually hide the nature of the entropy one is computing in the precise definition of the replica partition functions \cite{ss1}. We treat the case of a sphere computing the universal logarithmic term.

In order to compute the entanglement entropy we should consider the vacuum state in the algebra of gauge invariant operators. This later is generated by the curvature tensor which is gauge invariant at the linearized level. The vacuum is a Gaussian state in this algebra and we could apply EE formulas for Gaussian states in terms of the correlation functions and commutators of the Gaussian variables. However, due to the algebraic complexity of dealing with the four index curvature tensor and its commutators we will follow a different route which is physically equivalent and will allow us to simplify the computations considerably. We will use the metric perturbation tensor $h_{\mu\nu}$ as a generator of the algebra. This is not a physical variable and we need to fix the gauge. This is done taking into account the spherical symmetry of the problem by choosing a gauge that allow us to decouple the two radial modes for each angular momentum. However, as explained in \cite{Casini:2013rba},  while fixing the gauge converts a gauge field into a physical variable, the localization properties of these variables are very much gauge dependent. Hence we need to fix the gauge such that the gauge fixed $h_{\mu\nu}$ can be recovered from the curvature inside the region of interest for computing the EE. Otherwise, selecting the field and momentum variables in a region may compute the EE of an algebra unrelated to geometry. 

Since this gauge fixing procedure adapted to the region of interest has not been explicitly carried out in the literature before we find it instructive to see how this works in the simpler case of a Maxwell field first. We will treat the case of a Maxwell field between parallel planes in the next section and in a sphere in section \ref{maxwellsphere}. The results agree with \cite{Casini2016} where the algebra was defined directly in terms of the electric and magnetic fields instead of the gauge fixed vector potential $A_\mu$. In section \ref{gravitonplanos} we describe the theory of the linearized graviton and compute the EE between parallel planes. The case of a sphere is treated in section \ref{gravitonsphere} where we compute the logarithmic coefficient. We end with a discussion in section \ref{dis}, where we briefly compare with other results in the literature.

\section{Entanglement entropy of a Maxwell field between  parallel planes}
\begin{figure}[h]
    \centering
    \includegraphics[width=0.45\textwidth]{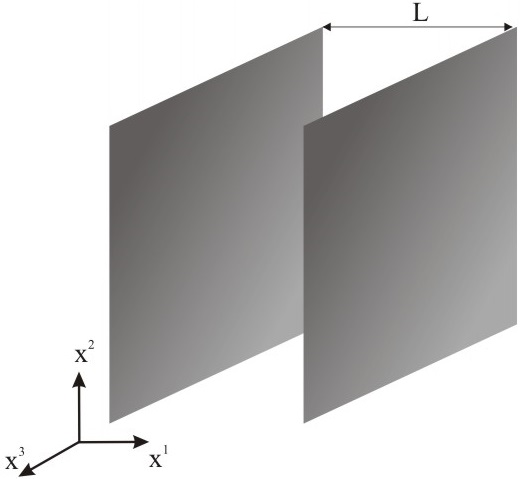}
    \caption{Two parallel planes with a separation of distance $L$ in the $x^1$ direction.}
    \label{paralelplates}
\end{figure}

Before studding the problem of linearized gravitons, we consider the simpler case of a free Maxwell field in $(3+1)$ dimensions given by the Lagrangian
\beq
L=-\frac{1}{4} \int d^3 x\, F_{\mu\nu} F^{\mu\nu}=\frac{1}{2} \int d^3 x \left[(\dot{\vec{A}}(\vec{x}) + \nabla A_0(\vec{x}))^2 - (\nabla \cross \vec{A}(\vec{x}))^2 \right]\,.
\label{MaxwellLagAmu}
\eeq
In this section we aim to obtain the EE associated with the region $\vv$  between two parallel planes separated by a distance $L$ (Figure \ref{paralelplates}). For a Cartesian coordinate system $\vec{x}=(x^1,x^2,x^3)$  the region $\vv$ is given by $\vv = \left\{{x=(x^1,x^2,x^3), 0<x^1<L}\right\}\, . $
In this context, it is particularly useful to write the field $A_\mu$ on a plane wave basis using the Fourier sum over the directions parallel to the plates. Assuming that the directions $x^2$ y $x^3$ are compactified to large sizes $R_2$ y $R_3$ with periodic boundary conditions, we are able to obtain
\beq
A_\mu(x^0,x^1,x^2,x^3)   = \sum_{\vec{k}} N e^{i\vec{k}\cdot\vec{x}}  A_\mu(x^0,x^1,k) ,
\label{ondasplanas1}
\eeq
where it is valid that $A_\mu^\dag(x^0,x^1,k)=A_\mu(x^0,x^1,-k)$ and $N$ is a constant that takes the value $\left[ \sqrt{2\pi R_2R_3}\right]^{-1} $. Moreover, the vector $\vec{k}$ can be expressed for $ n^2,n^3 = 0,\pm 1,\pm 2,...,\pm\infty$ as
\beq
\vec{k}=\left(0, \frac{2\pi n^2}{ R_2}, \frac{2\pi n^3}{R_3} \right) \, .
\eeq

The problem then decomposes into independent $(1+1)$-dimensional fields in the directions $x^0,x^1$, labeled by $\vec{k}$.  To study a fixed mode $\vec{k}$, without loss of generality, we can simplify the calculation by using a coordinate system adapted to $\vec{k}$, where $\hat{x}^2=\hat{k}$ and $\hat{x}^3 = \hat{x}^1 \cross \hat{k}$. In these coordinates, the expression of the each mode is 
\beq
 N e^{ikx_2} A_\mu (x_0,x_1, k) \, .
\label{AFourier}
\eeq

Considering the gauge freedom of the Maxwell field given by ${A'}_\mu \rightarrow A_\mu  + \partial_\mu \chi$ we also decompose $\chi$ in the plane wave basis. The mode corresponding to $\vec{k}$ writes
\beq
\chi (x_0,x_1, x_2, x_3) =N e^{i k x_2} \chi (x_0,x_1, k) \,.
\label{chichi}
\eeq
Then, a gauge transformation of a fix mode yields 
\bea
 A'_\mu (x_0, x_1, x_2, x_3) &=& N e^{ikx_2} \left[(A_0(x_0, x_1, k) + \dot{\chi} (x_0, x_1, k)) \hat{x}_0 +\right. \nonumber \\
&+&\left. (A_1 (x_0, x_1, k) + \partial_1 \chi (x_0, x_1, k))\hat{x}_1 \right. \nonumber \\
&+&\left.  (A_2 (x_0, x_1, k) + ik\chi (x_0, x_1, k))\hat{x}_2 +A_3 (x_0, x_1, k) \hat{x}_3 \right] \, . 
\label{maxwellgauge2}
\eea

In the light of (\ref{maxwellgauge2}), it is clear that we can fix $\chi$ in such a way that the field component parallel to $\hat{k}$ vanishes, $A_2=0$. 
With this choice we have  
\beq
F_{2\nu}=\partial_2 A_\nu - \partial_\nu A_2 = ikA_\nu \, .
\label{FmunuPP}
\eeq
In other words, in this gauge, and for each fixed mode $k$, $A_\mu$  can be expressed as a function of the gauge invariant tensor $F_{\mu\nu}$ in a way local in the coordinates $x^0,x^1$. This allows us to identify the algebra of gauge invariant operators $F_{\mu\nu}$ in between the parallel planes with the one of the quantized gauge fixed operators $A_\mu$.

To proceed we compute the Hamiltonian.  We must rewrite the Lagrangian (\ref{MaxwellLagAmu}) by using the expansion (\ref{AFourier}) under the proposed gauge condition. By doing so, we obtain for each mode the Lagrangian
\bea
\lag_k &=& 1/2 \left[\dot{A}^\dag_1\dot{A}_1+\dot{A}^\dag_3 \dot{A}_3 -k^2 {A}^\dag_1{A}_1 -k^2 {A}^\dag_3{A}_3 - \partial_1 A_3^\dag\partial_1 A_3  \right. \nonumber \\  
&& -\left. k^2 {A}^\dag_0{A}_0 - \partial_1 A_0^\dag\partial_1 A_0 - A_0^\dag\partial_1 \dot{A}_1   - \partial_1 \dot{A}^\dag_1  A_0  \right] \, .
\label{MaxwellPPlag}
\eea
The canonical momenta of the fields $A_1\,,A_1^\dag\, ,A_3\,,A_3^\dag$ are given by
\bea
\pi_1 = \frac{\partial \lag_k}{\partial \dot{A}_1 }= \frac{\dot{A}^\dag_1}{2}+ \frac{\partial_1 A^\dag_0}{2} \, \, , \, \, \pi_3 = \frac{\partial \lag_k}{\partial \dot{A}_3 }= \frac{\dot{A}^\dag_3}{2}\,,\\
\pi^\dag_1 = \frac{\partial \lag_k}{\partial \dot{A}^\dag_1 }= \frac{\dot{A}_1}{2}+ \frac{\partial_1 A_0}{2}  \, \, , \, \, \pi^\dag_3 = \frac{\partial \lag_k}{\partial \dot{A}^\dag_3 }= \frac{\dot{A}_3}{2}\,.
\eea
The Hamiltonian of the mode is then given by the Legendre transform
\bea
 \ham_k &=& \pi_1\dot{A}_1+\pi^\dag_1\dot{A}^\dag_1+\pi_3\dot{A}_3+\pi^\dag_3\dot{A}^\dag_3 - \lag_k = 2\pi_1^\dag \pi_1 + 2\pi_3^\dag \pi_3 + \frac{k^2}{2} A^\dag_1 A_1 + \nonumber \\
&+&  \frac{k^2}{2} A^\dag_3 A_3 +\frac{1}{2}\partial_1 A^\dag_3\partial_1 A_3  + \frac{k^2}{2} A^\dag_0 A_0 + A^\dag_0 \partial_1 \pi^\dag_1  +A_0 \partial_1 \pi_1 
\label{MaxwellPPham}
\eea
with the corresponding equal time commutation relations
\bea
\left[A_1(x_0,x_1,k),\pi_1({x}_0,{x'}_1,k)\right]=i \delta(x_1-{x'}_1) \, , \nonumber \\
\left[A_3(x_0,x_1,k),\pi_3({x}_0,{x'}_1,k)\right]=i \delta(x_1-{x'}_1)\, .
\label{MaxwellPPcom}
\eea
It is clear from (\ref{MaxwellPPham}) that the field $A_0$ does not posses his own dynamic and thus it can be treated as a Lagrange multiplier.  Differentiating, in order to obtain its equations of motion, we obtain the constraints
\be
\partial_1 \pi_1 = - \frac{k^2}{2} A^\dag_0 \,\, , \,\, \partial_1 \pi^\dag_1 = - \frac{k^2}{2} A_0 \, .
\label{MaxwellPPvinc}
\ee
Replacement of (\ref{MaxwellPPvinc}) in (\ref{MaxwellPPham}) gives
\be\ham_k= 2\pi_1^\dag \pi_1 + 2\pi_3^\dag \pi_3 + \frac{k^2}{2} A^\dag_1 A_1 + \frac{k^2}{2} A^\dag_3 A_3 +
+\frac{1}{2}\partial_1 A^\dag_3\partial_1 A_3  + \frac{2}{k^2} \partial_1 \pi_1 \partial_1 \pi^\dag_1\,.
\label{MaxwellPPham2}
\eeq
Making the identifications
\beq
\phi_1 = \frac{\sqrt{2}\pi_1}{|k|} \, , \, P_1=-\frac{|k|A_1}{\sqrt{2}}\,, 
\eeq
\beq
\phi_3 = \frac{A_3}{\sqrt{2}} \, , \, P_3=\sqrt{2} \pi_3\,,
\eeq
where $\phi_1, P_1$ and $\phi_3, P_3$ are pairs of canonically conjugate variables, 
the Hamiltonian writes 
\beq
\ham_k= P^\dag_1 P_1 + P^\dag_3 P_3 + \partial_1 \phi^\dag_1\partial_1 \phi_1 + \partial_1 \phi^\dag_3\partial_1 \phi_3 + k^2 \phi^\dag_1 \phi_1 + k^2 \phi^\dag_3 \phi_3\,.\label{217}
\eeq
This is exactly the Hamiltonian for the modes of two independent {\sl scalar} fields $\phi_1, \phi_3$ upon dimensional reduction (see for example \cite{Casini2016}). As a result, the algebra of gauge invariant operators for the gauge field and the vacuum expectation values inside the parallel planes are identical to the algebras and expectation values corresponding to two massless scalar fields inside the same region.    

To sum up, we conclude the EE of the Maxwell field associated with a region $\vv$ enclosed by two parallel planes is equivalent to the contribution of two independent scalar fields. In this way, we recover the known result obtained in \cite{Casini2016} by working with the gauge invariant electric and magnetic fields directly. The entropy turns out to be \cite{Casini:2005zv}
\be
S=c \frac{A}{\epsilon^2}- 2 \,k_s \frac{A}{L^2}\,,
\ee
where $A=R_2 R_3$ is the area of the planes, $\epsilon$ is a short distance cutoff, $c$ a non universal constant,  and $k_s$ is the universal coefficient corresponding to a scalar in this same geometry. This later can be computed with high precision from the knowledge of the one dimensional scalar entropy function \cite{Casini:2009sr}
\be
k_s=  0.0055351599...\,.\label{ks}
\ee
As we will now see, this exact identification of entropies between scalars and gauge fields does not hold for other regions.

\section{Entanglement entropy for a Maxwell field in the sphere}
\label{maxwellsphere}

We consider now the problem of a Maxwell field in the sphere, which also can be easily dimensionally reduced. Due to the spherical symmetry presented in this case we expand the field in question using scalar spherical harmonics for the $A_0$ component and vector spherical harmonics for $\vec{A}=(A_1,A_2,A_3)$. That is
\bea
A_0 &=& \sum_{lm} A^0_{lm} (t,r) \as (\theta, \phi)\,,\quad l=0,1,...,\infty\,, \quad  -l\leq m \leq l \, ,
\label{A0Ylm}\\
\vec{A} &=& \sum_{slm} A^s_{lm} (t,r) \av{s} (\theta, \phi) \,,\quad l=0,1,...,\infty\,, \quad  -l\leq m \leq l \,, \quad s=r,e,m \, ,
\label{MaxwellSbase}
\eea
where $\av{s}$ are the vector spherical harmonics defined by
\bea
\av{r}(\theta,\varphi) &=& \as(\theta,\varphi) \hat{r}\,, \qquad l \geq 0, \quad -l\leq m \leq l\,,
\label{Ylmr}\\
\av{e}(\theta,\varphi) &=& \frac{r \nabla \as(\theta,\varphi)}{\sqrt{l(l+1)}}\,,  \qquad l > 0, \quad -l\leq m \leq l\,,
\label{Ylme}\\
\av{m}(\theta,\varphi) &=& \frac{\vec{r} \times \nabla \as(\theta,\varphi)}{\sqrt{l(l+1)}}\,,  \qquad l > 0, \quad -l\leq m \leq l\,.
\label{Ylmm}
\eea

Considering the gauge transformations, it is useful to expand the function $\chi$ using scalar spherical harmonics as
\beq
\chi= \sum_{lm} \chi_{lm} (t,r) \as (\theta, \phi)\, .
\eeq
This gives the transformation law
\beq
\vec{A}'=\sum_{lm} \left( A_{lm}^{r}  + \partial_r \chi_{lm} \right) \av{r} + \left( A_{lm}^{e} + \frac{\chi_{lm}}{r} \right) \av{e} +  A_{lm}^{m} \av{m} \, .
\eeq

We see it is possible to fix $\chi_{lm}$ completely in such way that the coefficient $ {A'}_{lm}^{e}$ of the ``electric'' vector spherical harmonics $\av{e}$ is identically zero for each angular momentum. 
 This particular choice of gauge is convenient because of other reasons too. 
 It 
 allow us to write  for each mode
\beq 
F_{e\mu} =\left(e^\nu\partial_\nu\right) A_\mu  + \left( \partial_\mu e^\nu \right) A_\nu
\label{FmunuS}
\eeq
where $e^\mu$ is the unit vector in the direction of $\av{e}$,   $e^\mu A_\mu$ represents the electric component of the vector $A_\mu$ that vanishes under this particular choice of gauge, and $e^\mu \partial_\mu$ is the derivative in such direction. The expression (\ref{FmunuS}) shows that in this gauge we can recover $A_\mu$ on a sphere by the knowledge of the components  $F_{e\mu}$ of the gauge invariant field tensor in the same sphere. This is because derivatives in (\ref{FmunuS}) are tangential to the sphere. Therefore, even if the relation between the gauge fixed $A_\mu$ and $F_{\mu \nu}$ is non local in the angular directions, it maps the variables $A_\mu$ at fixed $r$ to physical variables with the same radius. This is a particular case of the general situation studied in \cite{Casini:2013rba} where it was shown that a gauge fixing that respects the localization of degrees of freedom in a region can be chosen such that $A_\mu$ vanishes on the boundary of the region in a direction parallel to the boundary itself. In the present example this direction is the one of the electric vector harmonics.  

From this point, we proceed in the same way as in the case of parallel planes. In particular, a useful writing  of the Lagrangian can be obtained by means of replacing (\ref{A0Ylm}) and (\ref{MaxwellSbase}) in (\ref{MaxwellLagAmu}), so as, by taking into consideration the orthonormality property of vector spherical harmonics, we get
\beq
L=\sum_{lm} \int_0^\infty dr \lag_{lm}\,.
\eeq
The Lagrangian $\lag_{lm}$ for $l\geq 1$ follows from direct computation using the properties of vector harmonics listed in appendix \ref{apa},
\bea
\lag_{lm}&=&1/2 \left[r^2 {\dot{A}_{l,m}^{r}}{\dot{A}_{l,-m}^{r}} + r^2 {\dot{A}_{l,m}^{m}}{\dot{A}_{l,-m}^{m}}-l(l+1){{A}_{l,m}^{r}}{{A}_{l,-m}^{r}}  \right.\nonumber\\
&&-\left. l(l+1){{A}_{l,m}^{m}}{{A}_{l,-m}^{m}} - \left|{{A}_{l,m}^{m}} + r\partial_r {{A}_{l,m}^{m}}\right|^2 + r^2 \partial_rA^0_{l,m}\partial_rA^0_{l,-m}+  l(l+1) A^0_{l,m}A^0_{l,-m}\right.\nonumber \\
&&- \left.  r^2 A^0_{l,m} \partial_r{\dot{A}_{l,-m}^{r}} -r^2 A^0_{l,-m} \partial_r{\dot{A}_{l,m}^{r}} + 2rA^0_{l,m}{\dot{A}_{l,-m}^{r}}+ 2rA^0_{l,-m}{\dot{A}_{l,m}^{r}}\right]\, .
\label{MaxwellSlag}
\eea
The Lagrangian density is independent of $m$, and to simplify the notation in the following we eliminate the index for $m$ in the variables and consider the real $m=0$ mode only, while we have to keep in mind that we will have $(2 l+1)$ identical contributions to the EE for each angular momentum $l$.   

The canonical conjugate momenta are defined by
\beq
\pi_{l}^r =\frac{\partial \lag_{l}}{\partial {\dot{A}_{l}^{r}}} = r^2\left({\dot{A}_{l}^{r}}+\partial_r A^0_l \right)\, , \quad \pi_{l}^m =\frac{\partial \lag_{l}}{\partial {\dot{A}_{l}^{m}}} = r^2{\dot{A}_{l}^{m}} \, .
\eeq
which  can be substituted in the Legendre transform
\beq
 \ham_{l}= \pi_{l}^r\dot{A}_{l}^{r} + \pi_{l}^m\dot{A}_{l}^{m} - \lag_{l} \, , 
\label{MaxwellSleg}
\eeq
in order to obtain the Hamiltonian
\bea 
\ham_{l} = \frac{\pi_{l}^r\pi_{l}^r}{2r^2}+ \frac{\pi_{l}^m\pi_{l}^m}{2r^2}+l(l+1){{A}_{l}^{r}}{{A}_{l}^{r}}+l(l+1){{A}_{l}^{m}}{{A}_{l}^{m}} \nonumber\\
+\left({{A}_{l}^{m}} + r\partial_r {{A}_{l}^{m}}\right)^2 -   \pi_{l}^r\partial_r A_{l}^0 - \frac{l(l+1)}{2} A_{l}^0 A_{l}^0\,.
\label{MaxwellSham}
\eea
The non trivial canonical commutation relations are given by
\beq
\left[{{A}_{l}^{r}}(t,r), {{\pi}_{l}^{r}}(t,r') \right]=\left[{{A}_{l}^{m}}(t,r), {{\pi}_{l}^{m}}(t,r') \right] = i \delta(r-r')\, .
\label{MaxwellScom}
\eeq
In addition, modes with different $l$ are independent to each other and their operators commute.

Again, $A_{l}^0$ is a Lagrange multiplier, allowing the derivation of the constraint
\beq
\partial_r \pi_{l}^r = l(l+1)A_{l}^0\,,
\label{MaxwellSvinc}
\eeq
which can be replaced in (\ref{MaxwellSham}) yielding
\bea
 \ham_{l} &=& \frac{1}{2}\left[\frac{\pi_{l}^r\pi_{l}^r}{r^2}+ \frac{\partial_r \pi_{l}^r\partial_r \pi_{l}^r}{l(l+1)} + {l(l+1)} A_{l}^r A_{l}^r\right] \nonumber \\
&+& \frac{1}{2}\left[\frac{\pi_{l}^m\pi_{l}^m}{r^2} + 2\left({{A}_{l}^{m}} + r\partial_r {{A}_{l}^{m}}\right)^2 + {l(l+1)} A_{l}^m A_{l}^m\right]
\label{MaxwellSham2} \, .
\eea
Lastly, the field and momentum variables can be rewritten as
\bea
\phi^r_{l} &=& \frac{\pi^r_{l}}{\sqrt{l(l+1)}}  \, , \,\,\, P^r_{l} =- \sqrt{l(l+1)} A^r_{l}\,,
\label{MaxwellScamp1}\\
\phi^m_{l} &=& r A^r_{l} \, , \quad\qquad\, P^m_{l} = \frac{\pi_{l}^m}{r}\,, 
\label{MaxwellScamp2}
\eea
and by applying (\ref{MaxwellScamp1}) and (\ref{MaxwellScamp2}) in (\ref{MaxwellSham2}) we reduce the Hamiltonian  to the one of two identical radial modes given by 
\bea
\ham_{l} &=& \frac{1}{2}\left[P^r_{l}P^r_{l}+ \partial_r \phi^r_{l}\partial_r \phi^r_{l} + \frac{l(l+1)}{r^2} \phi^r_{l}\phi^r_{l}\right]\nonumber\\
&+& \frac{1}{2}\left[P^m_{l}P^m_{l}+ \partial_r \phi^m_{l}\partial_r \phi^m_{l} + \frac{l(l+1)}{r^2} \phi^m_{l}\phi^m_{l}\right] \, ,\label{tyo}
\eea
with the standard commutation relations
\beq
\left[{{\phi}_{l}^{r}}(t,r), {{P}_{l}^{r}}(t,r') \right]=\left[{{\phi}_{l}^{m}}(t,r), {{P}_{l}^{m}}(t,r') \right] = i \delta(r-r') \, .
\label{MaxwellScom2}
\eeq

Each of these two identical modes has the same Hamiltonian as the one that results from the spherial reduction of a free massless scalar field \cite{Casini2016,Srednicki:1993im}. 

Eq. (\ref{MaxwellSlag}) does not apply to the zero angular momentum mode. This is simply because the electric (\ref{Ylme}) and magnetic (\ref{Ylmm}) spherical harmonics do not exist for $l=0$. For $l=0$ we get the simpler expression
\beq
\ham_{0} = \frac{\pi_{0}^r\pi_{0}^r}{2r^2}+ A_{0}^0 \partial_r\pi_{0}^r \, ,
\label{MaxwellSham3}
\eeq
where by replacing the constraint $\partial_r \pi_{l=0}^r =0$, obtained from the equations of motion of $A^0_{0}$, we get
\beq
\ham_{0} = \frac{\pi_{0}^r\pi_{0}^r}{2r^2}\, .
\label{MaxwellSham4}
\eeq
This means that the zero angular momentum mode does not have dynamics  and thus generates no contributions to the entropy.

Therefore, the EE of the Maxwell field on the sphere is equivalent to the one of two scalar fields where the $l=0$ mode has been subtraced. This result coincides with the one given in \cite{Casini2016}. 
 The entanglement entropy of a scalar in a sphere has a universal logarithmic term $-1/90 \log(R/\epsilon)$ \cite{scalar,scalar1,dowkeresferagauge,scalar2,scalar3}. The mode $l=0$ for the scalar (see the Hamiltonian (\ref{tyo}) for $l=0$) corresponds to a massless $d=2$ field in the $r>0$ half-line with Dirichlet boundary condition at the origin, and its universal logarithmic entropy is
  $1/6 \log(R/\epsilon)$ \cite{Casini2016,calcar}. 
The entropy of the Maxwell field in the sphere is then given by \cite{Casini2016,dowkeresferagauge} 
\be
S=c \frac{A}{\epsilon^2}-\frac{16}{45} \log(R/\epsilon)\,,
\ee
where the coefficient of the logarithmic term follows from $16/45=2\times 1/90+2\times 1/6$.

 Here we recover this result by working with the gauge variant field $A_\mu$ instead of using directly the electric and magnetic gauge invariant fields as in  \cite{Casini2016}. It is important to remark that another gauge choice which does not respect the locality on the sphere would have given completely different incorrect results for the sphere EE. 

\section{Entanglement entropy of linearized gravitons between parallel planes}
\label{gravitonplanos}

The free theory of a massless helicity $2$ particle can be described by a field $h_{\mu\nu}$. This field can be thought as describing metric perturbations $g_{\mu\nu}=\eta_{\mu\nu}+ h_{\mu\nu}$ with respect to the Minkowski metric $\eta_{\mu\nu}$. The field $h_{\mu\nu}$ obeys the linearized Einstein equations and the  Lagrangian that give these equations in absence of sources writes \cite{Ortin:2004ms}
\beq
\lag = -\partial_\mu h^{\mu\nu} \partial_\alpha h^\alpha_{\hphantom{\alpha}\nu} +\frac{1}{2}\partial^\alpha h_{\mu\nu} \partial_\alpha h^{\mu\nu} + \partial_\mu h^{\mu\nu} \partial_\nu h^\alpha_{\hphantom{\alpha}\alpha} - \frac{1}{2} \partial_\alpha h^\mu_{\hphantom{\mu}\mu} \partial^\alpha h^\nu_{\hphantom{\nu}\nu} 
\label{lag1} \, .
\eeq
The theory has a gauge invariance given by the transformation law 
\be
h'_{\mu\nu} = h_{\mu\nu} + \partial_\nu \xi_\mu + \partial_\mu \xi_\nu\,,\label{gg}
\ee
for arbitrary vector field $\xi_\mu$. This corresponds to the diffeomorphism invariance of the Einstein theory of gravity at the linearized level. 

The curvature is not gauge invariant in the non linear gravity theory but a gauge invariant operator corresponds to the linearized curvature tensor \cite{wee}
\beq
R_{\mu\nu\rho\sigma}= \frac{1}{2}\left[ \partial_\nu \partial_\rho h_{\mu\sigma}- \partial_\mu \partial_\rho h_{\nu\sigma}+ \partial_\mu \partial_\theta h_{\nu\rho}- \partial_\nu \partial_\sigma h_{\mu\rho}\right]\, .
\label{RDD}
\eeq
It is a simple exercise to show that it is indeed invariant under (\ref{gg}).\footnote{This corresponds to the fact that the curvature transforms linearly under changes of coordinates and it is already of linear order in $h_{\mu\nu}$.  Then further factors of the infinitesimal coordinate transformation must be second order.}  Therefore the theory of a helicity $2$ field in Minkowski space contains gauge invariant local operators, in contrast to what is expected in full quantum gravity. In consequence the EE is well defined, except for the usual issues about divergent terms. Let us study first the case of a region bounded by two parallel planes.

\subsection{Plane wave decomposition and gauge fixing}
 For the wall between parallel planes we resort to a plane wave decomposition of the fields analogous to (\ref{AFourier}). Now we write for the field of a mode with $\vec{k}=k \hat{x}_2$
\beq
h_{\mu\nu}(x_0,x_1,x_2,x_3)= N e^{ik x_2} h_{\mu\nu}(x_0,x_1,k)\, .
\label{ondasplanas2}
\eeq
The arbitrary gauge function $\xi$ can also be decomposed in modes. The mode with vector $\vec{k}$ writes 
\beq
\xi_{\mu} (x_0,x_1,x_2,x_3) =  N e^{ik x_2} \xi_{\mu}(x_0,x_1,k)\,.
\label{gauge2}
\eeq
Therefore, it can be easily observed that
\beq
h'_{\mu\nu}=\begin{bmatrix}{h_{00} + 2 \dot{\xi_{0}}}&{h_{01}+\dot{\xi_{1}}+\partial_1 \xi_0}&{h_{02}+\dot{\xi_{2}}+ik\xi_0}&{h_{03}+\dot{\xi_{3}}}\\{h_{01}+\dot{\xi_{1}}+\partial_1 \xi_0}&{h_{11}+2\partial_1\xi_1}&{h_{12}+\partial_1\xi_2+ik\xi_2}&{h_{13}+\partial_1 \xi_3 }\\{h_{02}+\dot{\xi_{2}}+ik\xi_0}&{h_{12}+\partial_1\xi_2+ik\xi_2}&{h_{22}+2ik\xi_2}&{h_{23}+ik\xi_3}\\{h_{03}+\dot{\xi_{3}}}&{h_{13}+\partial_1 \xi_3}&{h_{23}+ik\xi_3}&{h_{33}}\end{bmatrix}\,,
\label{matrixh}
\eeq
making it clear that the components $h'_{02}$, $h'_{20}$, $h'_{12}$, $h'_{21}$, $h'_{22}$, $h'_{23}$, y $h'_{32}$ can be fixed to zero if we use all the gauge freedom available. Now, all the components of $h'_{\mu\nu}$ that have zero contractions in the direction of $\hat{k}$ are fixed to zero, allowing us to write the following component of the Riemann tensor for each mode as
\beq
2R_{2\mu 2\nu} = h_{\nu 2},_{\mu 2} + h_{2\mu},_{2\nu} - h_{\nu\mu},_{22} - h_{22},_{\mu\nu} = - h_{\nu\mu},_{22} = k^2  h_{\nu\mu}\, .
\label{RiemannGauge}
\eeq
That is, under this particular choice of gauge, it is possible to write the field $h_{\nu\mu}$ in terms of the Riemann tensor in a local way in the $x_0$ and $x_1$ coordinates. This means that the algebra of the gauge fixed  degrees of freedom in $h_{\nu\mu}$ is the same as the algebra of the curvature tensor in between the planes.

\subsection{Lagrangian for each momentum}

The Lagrangian decomposes into independent modes for the different plane waves. Using the expansion (\ref{ondasplanas2}) and the gauge condition presented in the previous section, the Lagrangian density for the $(1+1)$ dimensional theory of the mode $k$ can be expressed as
\bea
\lag_k &=& \dot{h}_{13}\dot{h}_{13}^\dag - \frac{1}{2}(\dot{h}_{11}\dot{h}_{33}^\dag + \dot{h}_{33}\dot{h}_{11}^\dag) -k^2{h_{13}}{h_{13}}^\dag + \frac{k^2}{2}({h_{11}}{h_{33}}^\dag  +  {h_{33}}{h_{11}}^\dag)  \\
&&- \frac{k^2}{2}h_{00}({h_{11}}^\dag+{h_{33}}^\dag) - \frac{k^2}{2}({h_{11}}+{h_{33}})h_{00}^\dag-h_{01}\partial_1\dot{h}_{33}^\dag -\partial_1\dot{h}_{33}h_{01}^\dag +h_{03}\partial_1\dot{h}_{13}^\dag  \nonumber \\
  &+&\partial_1\dot{h}_{13}h_{03}^\dag +k^2 h_{03}h_{03}^\dag +k^2 h_{01}h_{01}^\dag + \partial_1 h_{03} \partial_1 h_{03}^\dag - \frac{1}{2}\partial_1 h_{33} \partial_1 h_{00}^\dag - \frac{1}{2}\partial_1 h_{00} \partial_1 h_{33}^\dag \,.\nonumber 
\label{lag4}
\eea

This Lagrangian contains two sets of independent fields and the problem can be split into two modes that will be treated separately. The first one contains the field $h_{13}$ and the Lagrange multiplier $h_{03}$ 
\beq
\lag_I = \dot{h}_{13}\dot{h}_{13}^\dag -k^2{h_{13}}{h_{13}}^\dag 
   -\partial_1 h_{03}\dot{h}_{13}^\dag -\dot{h}_{13}\partial_1 h_{03}^\dag +k^2 h_{03}h_{03}^\dag + \partial_1 h_{03} \partial_1 h_{03}^\dag\,,
\label{lag6}
\eeq
and the second one containing the fields $h_{11}$ y $h_{33}$ and the multipliers $h_{01}$ y $h_{00}$
\bea 
\lag_{II} &=& - \frac{1}{2}(\dot{h}_{11}\dot{h}_{33}^\dag + \dot{h}_{33}\dot{h}_{11}^\dag) + \frac{k^2}{2}({h_{11}}{h_{33}}^\dag  +  {h_{33}}{h_{11}}^\dag) + \partial_1 h_{01}\dot{h}_{33}^\dag +\dot{h}_{33}\partial_1 h_{01}^\dag  \\
  &&- \frac{k^2}{2}h_{00}({h_{11}}^\dag+{h_{33}}^\dag) - \frac{k^2}{2}({h_{11}}+{h_{33}})h_{00}^\dag +k^2 h_{01}h_{01}^\dag  - \frac{1}{2}\partial_1 h_{33} \partial_1 h_{00}^\dag - \frac{1}{2}\partial_1 h_{00} \partial_1 h_{33}^\dag\,. \nonumber
\label{lag7}
\eea
Therefore, the total Lagrangian is given by sum over modes 
\beq
  L = \sum_k \int_0^\infty dx_1 \left( \lag_I + \lag_{II}  \right)\,.
\label{lag8}
\eeq
\subsection{Hamiltonian of the mode I}
The momenta $\pi_{13}$, $\pi_{13}^\dag$ corresponding to the Lagrangian (\ref{lag6}) are 
\beq
\pi_{13}=\frac{\partial \lag_I}{\partial \dot{h}_{13}}= \dot{h}_{13}^\dag - \partial_1 h_{03}^\dag \quad , \quad \pi_{13}^\dag =\frac{\partial \lag_I}{\partial \dot{h}_{13}^\dag}=\dot{h}_{13}- \partial_1 h_{03} \, ,
\label{mom}
\eeq 
and the corresponding Hamiltonian is
\bea
\ham_I &=& \pi_{13}\dot{h}_{13} + \pi_{13}^\dag\dot{h}_{13}^\dag - \lag_I =\nonumber \\
&=& \pi_{13}\pi_{13}^\dag  + k^2{h_{13}}{h_{13}}^\dag 
   - h_{03} \partial_1 \pi_{13} - h_{03}^\dag \partial_1 \pi_{13}^\dag - k^2 h_{03}h_{03}^\dag \, . 
\label{ham1}
\eea
A constraint equation can be derived by computing the equation of motion of $h_{03}$
\beq
h_{03} = \frac{\partial_1 \pi_{13}^\dag}{k^2} \quad , \quad  h_{03}^\dag = \frac{\partial_1 \pi_{13}}{k^2} \, ,
\label{constrains}
\eeq
which can be replaced in (\ref{ham1}) to obtain
\beq
\ham_I = \pi_{13}\pi_{13}^\dag  + k^2{h_{13}}{h_{13}}^\dag + \frac{\partial_1 \pi_{13}\partial_1 \pi_{13}^\dag}{k^2} \, .
\label{ham2}
\eeq
In order to rewrite the Hamiltonian (\ref{ham2}) as the one associated with a complex scalar (for each pair $\vec{k},-\vec{k}$) is convenient to define
\beq
\phi_{I} = \frac{\pi_{13}}{|k|} \quad , \quad P_{I} = -|k|h_{13} \,,
\label{replace1}
\eeq
\beq
 \phi_{I}^\dag = \frac{\pi_{13}^\dag}{|k|} \quad , \quad P_{I}^\dag = -|k|h_{13}^\dag\,.
\label{replace2}
\eeq
In this way (\ref{ham2}) gets 
\beq
\ham_I = P_{1}P_{1}^\dag  + k^2{\phi_{1}}{\phi_{1}}^\dag + {\partial_1 \phi_{1}\partial_1 \phi_{1}^\dag} \, .
\label{ham3}
\eeq
Moreover, by canonical quantization of the field ${h_{13}}$ in (\ref{ham1}), it is clear that the replacements (\ref{replace1}) and (\ref{replace2}) give  $P_{1}$ and $\phi_1$ as canonically conjugate variables.
\subsection{Hamiltonian of the mode II}
Analyzing the dynamic of the fields $h_{00}$  y $h_{00}^\dag$ in the Lagrangian (\ref{lag7}) is evident that both play the role of Lagrange multipliers, giving rise to the constraints
\beq
\partial_1\partial_1 h_{33} = k^2({h_{11}}+{h_{33}}) \, , \quad
\partial_1\partial_1 h_{33}^\dag = k^2({h_{11}}^\dag+{h_{33}}^\dag) \, .
\label{cons1}
\eeq
If we replace (\ref{cons1}) in (\ref{lag7}) it is possible to eliminate the field $h_{11}$ from the Lagrangian, obtaining
\bea
\lag_{II} &=&  \dot{h}_{33}\dot{h}_{33}^\dag - k^2 {h_{33}}{h_{33}}^\dag + \frac{\partial_1\dot{h}_{33}\partial_1\dot{h}_{33}^\dag}{k^2} \nonumber \\
 &&- \partial_1{h_{33}}\partial_1{h_{33}}^\dag - h_{01}\partial_1\dot{h}_{33}^\dag - \partial_1\dot{h}_{33} h_{01}^\dag +k^2 h_{01}h_{01}^\dag  \, . 
\label{lag111}
\eea
The substitution produced a higher derivative term  $k^{-2}\partial_1\dot{h_{33}}\partial_1\dot{h_{33}}^\dag$. However, this problem disappear  when we use the constraint equation related to the equation for the fields  $h_{01}$  y $h_{01}^\dag$ that also work as Lagrange multipliers. Indeed, we obtain
\beq
 h_{01} = \frac{\partial_1\dot{h}_{33}}{k^2}   \quad , \quad
 h_{01}^\dag = \frac{\partial_1\dot{h}_{33}^\dag}{k^2}   \, .
\label{cons21}
\eeq
Applying (\ref{cons21}) in (\ref{lag111}) the Lagrangian gets reduced to
\beq
 \lag_{II} =  \dot{h}_{33}\dot{h}_{33}^\dag - k^2 {h_{33}}{h_{33}}^\dag  - \partial_1{h_{33}}\partial_1{h_{33}}^\dag 
\label{lag13} \, .
\eeq
The canonical momenta associated to the complex field variables ${h_{33}}$ and ${h_{33}}^\dag$ are
\beq
 \pi_{33}=\frac{\partial\lag_{II}}{\partial \dot{h}_{33}} = \dot{h}_{33}^\dag \quad , \quad \pi_{33}^\dag=\frac{\partial\lag_{II}}{\partial \dot{h}_{33}^\dag} =\dot{h}_{33} \, ,
\label{mom5}
\eeq
and the corresponding  Hamiltonian gets the same form as (\ref{ham3}),
\beq
\ham_{II} = P_{II}P_{II}^\dag  + k^2{\phi_{II}}{\phi_{II}}^\dag + {\partial_1 \phi_{II}\partial_1 \phi_{II}^\dag} \, ,
\label{ham6}
\eeq
where we have introduced the trivial notation change 
\bea
\phi_{II} &=& {h_{33}} \quad , \quad \,\, P_{II} = \pi_{13}\,, 
\label{replace3}\\
  \phi_{II}^\dag &=& {h_{33}}^\dag \quad , \quad P_{II}^\dag = \pi_{13}^\dag \,.
\label{replace4}
\eea
These variables also obey canonical commutation relations. 
\subsection{Entanglement Entropy}
The Hamiltonians for the two modes (\ref{ham3}) and (\ref{ham6}) are equivalent to the ones of a dimensionally reduced scalar field, and to the two modes (\ref{217}) for the Maxwell field. Therefore, we are allowed to conclude that the EE of linearized gravitons for the region enclosed between two parallel planes is equivalent to the one of two scalar fields or one Maxwell field. The universal coefficients will be the same in the three cases. We get again  
\be
S=c \frac{A}{\epsilon^2}- 2 \,k_s \frac{A}{L^2}\,,
\ee
with $k_s$ given by (\ref{ks}) \cite{Casini:2005zv}.

\section{Entanglement entropy of linearized gravitons in a sphere}
\label{gravitonsphere}

In this section we treat the case of gravitons inside a sphere. We first introduce the tensor spherical harmonics which we use to decompose $h_{\mu\nu}$ in spherical coordinates. We also decompose the gauge transformations and choose a generic gauge adapted to the spherical symmetry that depends on three arbitrary constants. Then we expand the Lagrangian in terms of the gauge fixed field to get two independent radial modes for each $l,m$. The gauge choice is further refined to allow the simplification of the the mode Hamiltonians and to ensure locality in the radial direction in the relation between the gauge fixed field and the curvature tensor. We get a system of modes that are equivalent to the scalar spherical modes except that the $l=0,1$ modes are absent. Finally, we compute the entanglement entropy.     

\subsection{Tensor spherical harmonics}
The tensor spherical harmonics  are a further generalization of the concept of scalar and vector spherical harmonics that can be used as a basis for the space of symmetric tensors (of dimension six). An arbitrary symmetric tensor field $X$  can be expanded in polar coordinates as follows
\beq
X= \sum_{Jslm} X_{lm}^{Js}(r) T_{lm}^{Js}(\theta,\phi)\,,\,\, l=0,1,...,\infty\,,\,\, m=0,\pm 1,..., \pm l \, , \,\, Js=0l,0t,1e,1m,2e,2m\, ,  
\label{tensorexpand}
\eeq
where the tensor spherical harmonics $T_{lm}^{Js}$ are given by (see for example \cite{Compere2018,Thorne1980})
\bea
T_{lm}^{0l}&=& \hat{r} \otimes \hat{r} Y_{lm} \, , \,\,\,\qquad\qquad\qquad\qquad T_{lm}^{0t}= \frac{1}{\sqrt{2}} \left(\delta - \hat{r} \otimes \hat{r} \right) Y_{lm} \, ,\nonumber \\
T_{lm}^{1e}&=& \sqrt{\frac{2}{l(l+1)}}r  \left[\hat{r} \otimes \nabla Y_{lm} \right]^S \, , \,\,\quad \,\,\, T_{lm}^{1m}= \sqrt{\frac{2}{l(l+1)}}  \left[\hat{r} \otimes \overline{r} \cross \nabla Y_{lm} \right]^S\, , \label{22} \\
T_{lm}^{2e}&=& \sqrt{2\frac{(l-2)!}{(l+2)!}}  \left[r^2 \nabla\nabla Y_{lm} \right]^{STT} \, , \,\,\,\, T_{lm}^{2m}= \sqrt{2\frac{(l-2)!}{(l+2)!}} \left[r \nabla \left(\overline{r} \cross \nabla Y_{lm}\right) \right]^{STT}  \, . \nonumber
\eea
The spherical harmonics of spin $J=0$ are defined for $l\geq 0$, the ones related to spin $J=1$ for $l\geq 1$ and the in the case of spin $J=2$ for $l\geq 2$. In the notation of the equations (\ref{22}) the symbol $\delta$ means the identity tensor $\delta_{ij}$.   Additionally, the superscript $S$  means taking the symmetric part, and  $TT$ the traceless part transverse to $\hat{r}$. For an arbitrary tensor $X_{ij}$ this later is given by the following expression
\beq
X_{ij}^{TT}=\left(\delta_{ik} - \hat{r}_i\hat{r}_k \right)\left(\delta_{jn} - \hat{r}_j\hat{r}_n \right)X_{kn}-\frac{1}{2}\left(\delta_{ij} - \hat{r}_i\hat{r}_j \right)\left[\left(\delta_{kn} - \hat{r}_k\hat{r}_n \right)X_{nk}\right] \, .
\label{trasversetraceless}
\eeq 

It will be useful to have a relation between tensor and vector spherical harmonics. This relation can be expressed as
\bea
T_{lm}^{0l}&=& \left[ \hat{r} \otimes \overline{Y}_{lm}^r\right]^{S} \quad , \qquad \,\, T_{lm}^{0t} = \frac{1}{\sqrt{2}} \left(\delta  Y_{lm} - \hat{r} \otimes \overline{Y}_{lm}^r \right) \, , \nonumber \\
T_{lm}^{1e}&=& \sqrt{2} \left[\hat{r} \otimes \overline{Y}_{lm}^e  \right]^S \quad , \quad T_{lm}^{1m}= \sqrt{2}\left[\hat{r} \otimes \overline{Y}_{lm}^m \right]^S \, , \nonumber \\
T_{lm}^{2e} &=&   \sqrt{\frac{2}{(l-1)(l+2)}} \left\{ \left[r \nabla \overline{Y}_{lm}^e  \right]^{S} + \frac{1}{\sqrt{2}} \at{1e} + \sqrt{\frac{l(l+1)}{2}} \at{0t} \right\} \,, \label{33} \\
T_{lm}^{2m}&=& \sqrt{\frac{2}{(l-1)(l+2)}}  \left\{ \left[r \nabla \overline{Y}_{lm}^m \right]^{S} + \frac{1}{\sqrt{2}} \at{1m} \right\} \, .  \nonumber 
\eea

Further properties of the tensor spherical harmonics are listed in appendix \ref{apb}. 

\subsection{Decomposition of the spin 2 field in spherical harmonics}
To make the process of computing the EE easier it will be useful to decompose the field $h_{\mu\nu}$ and the gauge arbitrary function $\xi_\mu$ in different basis with spherical symmetry. For this purpose  we introduce  the notation

\beq
h_T=\begin{bmatrix}{h_{11}}&{h_{12}}&{h_{13}}\\{h_{21}}&{h_{22}}&{h_{23}}\\{h_{31}}&{h_{32}}&{h_{33}}\end{bmatrix}\, \, , \, \, h_V=\begin{bmatrix}{h_{01}}\\{h_{02}}\\{h_{03}}\end{bmatrix} \, \, , \, \, h_S={h}_{00}  \, \, , \, \, \xi_V=\begin{bmatrix}{\xi_1}\\{\xi_2}\\{\xi_3}\end{bmatrix} \, \, , \, \, \xi_S=\xi_0 \, .
\label{descomponemeelespacio}
\eeq

Firstly, we will work with the space-like part of the problem by expanding $h_T$ in tensor spherical harmonics and $\xi_V$ in vector spherical harmonics. Then we will study the time-like part by using vector spherical harmonics for $h_V$ and scalar spherical harmonics for $\xi_S$ and $h_S$.\footnote{See \cite{Compere2018,ReggeWheeler1957} for a different but somewhat analogous treatment of gravitons in spherical coordinates.}

\subsection{Gauge fixing for the space-like components}
As we just mention, $h_T$ y $\xi_V$ will be expanded using tensor and vector spherical harmonics respectively in the following way
\beq
h_{T}=\sum_{Jslm} h_{lm}^{Js}(t,r) T_{lm}^{Js}(\theta,\varphi) \quad , \quad \xi_{V}=\sum_{slm} \xi_{lm}^{s}(t,r) \overline{Y}_{lm}^{s}(\theta,\varphi) \, .
\label{hijexpand}
\eeq
On the other hand, the gauge freedom of linear gravity can be expressed in this notation as
\beq
h'_T= h_T + \nabla \xi_V + \left[\nabla \xi_V\right]^T = h_T + 2 \left[\nabla \xi_V\right]^S \, .
\label{ge1}
\eeq
The combination of (\ref{hijexpand}) and (\ref{ge1}) gives
\beq
h'_T = \sum_{Jslm} h_{lm}^{Js} T_{lm}^{Js} + 2 \sum_{slm} \left[\xi_{lm}^{s} \nabla \overline{Y}_{lm}^{s} +  \overline{Y}_{lm}^{s} \otimes \partial_r \xi_{lm}^{s}
\hat{r}\right]^S
\label{ge4} \, .
\eeq
By computing $\xi_{lm}^{s} \nabla \overline{Y}_{lm}^{s} +  \overline{Y}_{lm}^{s} \otimes \partial_r \xi_{lm}^{s}\hat{r}$ using the properties of vector and tensor spherical harmonics (appendices \ref{apa} and \ref{apb}), for $s=r,\, e,\, m$ separately, and then adding up these contributions we get
\bea
&& h'_T = \sum_{lm} \left( h_{lm}^{0l}+2\partial_r \xi_{lm}^{r} \right) T^{0l}_{lm} + \left( h_{lm}^{0t}  + \frac{2\sqrt{2}}{r} \xi_{lm}^r - \frac{\sqrt{2l(l+1)}}{r} \xi_{lm}^e\right) T_{lm}^{0t} \nonumber\\
&&+\left( h_{lm}^{1e} +\frac{\sqrt{2l(l+1)}}{r}\xi_{lm}^{r} + \sqrt{2}\partial_r \xi_{lm}^{e}  -\frac{\sqrt{2}}{r}\xi_{lm}^{e}\right) T^{1e}_{lm} + \left( h_{lm}^{1m} + \sqrt{2}\partial_r \xi_{lm}^{m} - \frac{\sqrt{2}}{r}\xi_{lm}^{m}  \right) T^{1m}_{lm} \nonumber\\
 &&+ \left( h_{lm}^{2e} + \frac{\sqrt{2(l-1)(l+2)}}{r}\xi_{lm}^{e}\right) T^{2e}_{lm}  + \left( h_{lm}^{2m} + \frac{\sqrt{2(l-1)(l+2)}}{r}\xi_{lm}^{m} \right) T^{2m}_{lm} \, .
\label{ge13}
\eea
This particular case differs from the ones studied earlier because there are many possible reasonable choices of gauge fixing for the spherical waves, but, not all of them will allow us to calculate the EE that corresponds to the spherical boundary or allow us to decouple the two dynamical modes for each $lm$. More specifically, it can be seen that
\begin{itemize}
    \item Fixing $ \xi_r $ allows us to cancel the components that are parallel to  $ \at{0t} $ or $ \at{1e} $ or to a linear combination of them.
    \item  Fixing $ \xi_e $ allows us to cancel the components that are parallel to $ \at{0t} $ or $ \at{2e} $ or  to a linear combination of them.
    \item  Fixing $ \xi_m $ allows us to cancel the components that are parallel to $ \at{2m} $.
\end{itemize}
\newpage

For now, we will use the freedom related to $\xi_m$ to cancel the 'electric-magnetic' components, meaning that we take ${h'}_{lm}^{2m}=0$ for all $l$ y $m$. Understanding the gauge fixing of $\xi_r$ y $\xi_e$ that is the correct one for our purposes is not simple at this stage. Because of that we choose to set to zero just some arbitrary linear combination of $\at{0t}$, $\at{1e}$ y $\at{2e}$ to be further determined in what follows. There is only one resulting degree of freedom that we call $\h{te}$ that is associated with a linear combination of these tensors given by some undetermined coefficients. More
formally, we fix the gauge such that
\beq
h_T = \sum_{lm} {h}^{0l}_{lm} \at{0l} + {h}^{te}_{lm} \left(\alpha \at{0t} +\beta \at{1e} +\gamma \at{2e} \right)  + {h}^{1m}_{lm} \at{1m}\,,
\label{ge15}
\eeq
where $\alpha$, $\beta$ and $\gamma$ are constants.
\subsection{Gauge fixing for the time-like components}
In order to  fix the gauge of the time-like part we will write the vector  $h_V$ and the scalar $\xi_S$ as
\beq
h_V=\sum_{slm} h_{lm}^{0s}(t,r) \overline{Y}_{lm}^{s}(\theta,\varphi)
\quad,\quad \xi_{S}=\sum_{lm} \xi^0_{lm}(t,r) {Y}_{lm}(\theta,\varphi) \, .
\label{xioexpand}
\eeq
For each component of $h_V$ we have ${h'}_{0i}=h_{0i}+\partial_0 \xi_i + \partial_i \xi_0$ or more conveniently ${h'}_V=h_V + {\dot{\xi}}_V + \nabla \xi_S $. By replacing with (\ref{xioexpand}) we get
\beq
{h'}_V=\sum_{lm} \left( h_{lm}^{0r} + \dot{\xi}^{r}_{lm} + \partial_r \xi^0_{lm} \right) \overline{Y}_{lm}^{r} + \left( h_{lm}^{0e} + \dot{\xi}^{e}_{lm} + \frac{\xi^0_{lm}}{r} \right) \overline{Y}_{lm}^{e} + \left( h_{lm}^{0m} + \dot{\xi}^{m}_{lm}\right) \overline{Y}_{lm}^{m} \, .
\label{gt3}
\eeq
Thus, in analogy with the case of the Maxwell field can fix $\xi_0$ in such way that $ {h'}_{lm}^{0e}$  is zero for each $lm$, obtaining the expansion
\beq
{h}_T=\sum_{lm}  {h}_{lm}^{0r} \overline{Y}_{lm}^{r} + {h}_{lm}^{0m} \overline{Y}_{lm}^{m}\, .
\label{gt4}
\eeq

\subsection{Lagrangian for each angular momentum}
The starting point is the Lagrangian (\ref{lag1}). Using the decomposition of the field $h_{\mu\nu}$ given in (\ref{descomponemeelespacio}) in terms of spatial and temporal components we obtain
\bea
\lag &=& \left. \frac{1}{2}\left(\dot{h}_T\cdot\cdot\dot{h}_T-\traza{\dot{h}_T}\traza{\dot{h}_T}\right)+\frac{1}{2}\left(\nabla^2 h_T \cdot\cdot  h_T + \nabla \traza{h_T} \cdot \nabla \traza{h_T}\right) \right. \nonumber \\
&&+ \left. \left(\nabla \cdot h_T\right) \cdot \left[\left(\nabla \cdot h_T\right)- \nabla \traza{h_T} \right]+ \nabla h_S \left[ (\nabla \cdot h_T) - \nabla \traza{h_T}  \right] \right. \\
&&- \left. 2\dot{h}_V\cdot\left[ (\nabla \cdot h_T) - \nabla \traza{h_T}  \right]-(\nabla \cdot h_V) \cdot (\nabla \cdot h_V)-\nabla^2 h_V \cdot  h_V \right. \,.\nonumber
\label{lag2ss2}
\eea
In this matricial notation a single dot means the contraction of a one index for each tensor and two dots the contraction of the two sets of indices of the two symmetric tensors involved in the product. 
The full Lagrangian is given  by
\beq
L=\int_\rrealt d^3 \overline{x}\, \lag = \int_0^\infty dr \, r^2 \left(\int  d\Omega \, \lag \right) \, .
\label{GravSllagtot}
\eeq

Replacing the expressions (\ref{ge15}) and (\ref{gt4}) in (\ref{lag2ss2}) and taking into account the properties of spherical harmonics (appendices \ref{apa} and \ref{apb}) it turns out we can rewrite the Lagrangian for independent modes for each $l$ y $m$ 
\beq
L=\sum_{lm} \int_0^\infty dr \, \left(\lag^I_{lm} + \lag^{II}_{lm} \right) \, ,
\eeq
where the $\lag_{lm}^I$ contains the variables $\h{1m}$ and $\h{0m}$ and $\lag_{lm}^{II} $ involves the fields $\h{0l}$ and $\h{te}$ together with the Lagrange multipliers $\h{0r}$ and $\h{00}$.
The Lagrangians for the modes are independent of $m$, and it is clear that we will have $(2l+1)$ equal contributions for each $l$. Accordingly we will suppress the index $m$. After a long but straightforward calculation using the properties listed in appendices \ref{apa} and \ref{apb}, the Lagrangian corresponding to the mode $I$ (and $l\ge 2$) gets
\bea
\lag_{l}^I &=& \frac{r^2}{2} \dot{h}^{1m}_l\dot{h}^{1m}_l  -\frac{(l-1)(l+2)}{2}\h{1m}\h{1m} +r^2\drh{0m}\drh{0m} \nonumber \\
&&+ l(l+1)\h{0m}\h{0m} +\sqrt{2}\dot{h}^{1m}_l\left( r\h{0m} - r^2\drh{0m}\right) \,,
\label{GravSmodI}
\eea
and the one associated with mode $II$ is
\bea
\lag_{l}^{II} &=& \frac{r^2}{2} \left(\beta^2-\alpha^2+\gamma^2\right)\dot{h}^{te}_l\dot{h}^{te}_l-\sqrt{2}r^2\alpha \dot{h}^{0l}_l\dot{h}^{te}_l+ \frac{r^2}{2}\left(\alpha^2-\gamma^2\right)\drh{te}\drh{te} \nonumber \\
&+&\sqrt{2}\alpha r \h{te}\drh{0l} +\h{0l}\h{0l}+\left(\beta^2-\frac{\sqrt{l(l+1)}}{2}\alpha\beta-\frac{\sqrt{(l-1)(l+2)}}{2}\beta\gamma\right) \h{te}\h{te}\nonumber \\
&+&\sqrt{2}\left( \frac{l(l+1)}{2}\alpha - \sqrt{l(l+1)}\beta + \frac{\sqrt{(l-1)l(l+1)(l+2)}}{2}\gamma \right)\h{0l}\h{te}+l(l+1)\h{0r}\h{0r} \nonumber \\
&+&\h{0r}\left[ 4r\dot{h}^{0l}_l -2\sqrt{2}\alpha r^2 \partial_r \dot{h}^{te}_l - \sqrt{2}\left(2 \alpha + \sqrt{l(l+1)}\beta\right)r \dot{h}^{te}_l\right]+\h{00} \left[-2r\drh{0l} \right. \nonumber \\ 
&-&\left. (l(l+1)+2)\h{0l} +\sqrt{2}\alpha r^2 \partial_r\drh{te}+\sqrt{2}\left(3\alpha+\sqrt{l(l+1)}\beta \right) r \drh{te}\right. \nonumber \\
&+& \left. \frac{1}{\sqrt{2}}\left(-(l-1)(l+2)\alpha+ 4 \sqrt{l(l+1)}\beta-\sqrt{(l-1)l(l+1)(l+2)}\gamma\right)\h{te} \right] \, . 
\label{GravSmodII}
\eea
In the same way as for the case of parallel planes, we will study the modes $I$ and $II$ separately trying to reduce them to scalar fields for $l\geq 2$. Then we will present the particular cases $l=0$ and $l=1$.

\subsection{Hamiltonian of mode I for \texorpdfstring{$l\geq 2$}{Lg}}
From equation (\ref{GravSmodI}) it can be seen clearly that $\h{0m}$ has no dynamic, thus we get the following constraint
\beq
-2r^2\partial_r\drh{0m}-4r\drh{0m}+2l(l+1)\h{0m}+\sqrt{2}r^2\dot{\drh{1m}}+3\sqrt{2}r\dot{h}^{1m}_l=0\,.
\label{GravSnoloc}
\eeq
In an analogy with the parallel planes case in equation (\ref{GravSnoloc}), this expression cannot be solved algebraically but the constraint can be implemented by first computing the Hamiltonian. The momenta are given by
\beq
\pi_{l}^{1m}= \frac{\partial \lag^I_{l}}{\partial \dot{h}^{1m}_l } = r^2 \dot{h}^{1m}_l + \sqrt{2}\left( r\h{0m} - r^2\drh{0m}\right) \, .
\label{smod11}
\eeq
From equations (\ref{GravSmodI}) and (\ref{smod11}) we compute
\bea
\ham^I_{l} &=& \pi_{l}^{1m}\dot{h}^{1m}_l-\lag^I_{lm} =\frac{\pi_{lm}^{1m}\pi_{lm}^{1m}}{2r^2}  + \frac{(l-1)(l+2)}{2} \h{1m}\h{1m}  \nonumber \\
&-& (l-1)(l+2)\h{0m}\h{0m}-\sqrt{2}\h{0m}\left(\partial_r \pi_{lm}^{1m} + \frac{\pi_{lm}^{1m}}{r} \right) \, .
\label{smod14}
\eea
Now, by working with $\h{0m}$ as a Lagrange multiplier in (\ref{smod14}) the following constraint appears
\beq
-2(l-1)(l+2)\h{0m}-\sqrt{2}\left(\partial_r \pi_{lm}^{1m} + \frac{\pi_{lm}^{1m}}{r} \right)=0
\label{smod15} \, .
\eeq
Replacing (\ref{smod15}) in (\ref{smod14}) gives for $l\geq 2$ 
\beq 
\ham^I_{l} = \frac{l(l+1)}{2r^2}\frac{\pi_{l}^{1m}\pi_{l}^{1m}}{(l-1)(l+2)}  + \frac{1}{2}\frac{\partial_r\pi_{l}^{1m}\partial_r\pi_{l}^{1m}}{(l-1)(l+2)} + \frac{1}{2}(l-1)(l+2)\h{1m}\h{1m} \, .
\label{smod17}
\eeq
and by redefining the variables
\beq
\phi^I_{l}=\frac{\pi_{l}^{1m}}{\sqrt{(l-1)(l+2)}}\,, \qquad P^I_{l}=-\sqrt{(l-1)(l+2)}\h{1m} \, ,
\label{smod18}
\eeq
we reduce (\ref{smod17}) to the Hamiltonian of a free scalar in the sphere
\beq 
\ham^I_{l} = \frac{1}{2}\left(P^I_{l}P^I_{l}  + \partial_r\phi^I_{l}\partial_r\phi^I_{l} + \frac{l(l+1)}{r^2}\phi^I_{l}\phi^I_{l} \right)\,.
\label{GravSSI}
\eeq
The canonical commutation relations
\beq
\left[P^I_{l}(t,r), \phi^I_{l}(t,r')\right]=i \delta(r-r') \, ,
\label{GravScomI}
\eeq
follow from 
\beq
\left[\pi^{1m}_{l}(t,r), h^{1m}_{l}(t,r')\right]=i \delta(r-r')\, .
\eeq
\subsection{Hamiltonian of mode II for \texorpdfstring{$l\geq 2$}{Lg}}
For the mode II we have the Lagrangian (\ref{GravSmodII}), where working out the equations of motion of $\h{00}$ yields the constraint
\bea
&&-2r\drh{0l}  -(l(l+1)+2)\h{0l} +\sqrt{2}\left(3\alpha+\sqrt{l(l+1)}\beta \right) r \drh{te}+\sqrt{2}\alpha r^2 \partial_r\drh{te} \nonumber \\
&&+ \frac{1}{\sqrt{2}}\left(-(l-1)(l+2)\alpha+ 4 \sqrt{l(l+1)}\beta-\sqrt{(l-1)l(l+1)(l+2)}\gamma\right)\h{te} =0 \, .
\label{GravSvinc}
\eea
Taking into account that (\ref{GravSvinc}) gives rise to non local terms (that can not be eliminated by the same means used for mode I), we are led to propose a particular gauge fixing such that
\beq
\h{0l}=a \h{te} + br\drh{te}\,,
\label{GravSreplace}
\eeq
with $a$ y $b$ constants that will be fixed to satisfy (\ref{GravSvinc}). Indeed, by replacing (\ref{GravSreplace}) in (\ref{GravSvinc}) we get
\bea
&&\sqrt{2}\left(\alpha - \sqrt{2} b \right) r^2 \partial_r \drh{te} + \left(3\sqrt{2}\alpha + \sqrt{2l(l+1)}\beta -(l(l+1)+4)b-2a\right)r\drh{te} \nonumber \\
&&\frac{1}{\sqrt{2}}\left(-(l-1)(l+2)\alpha + 4\sqrt{l(l+1)} \beta - \sqrt{(l-1)l(l+1)(l+2)}\gamma \right.\nonumber\\
&&\hspace{8cm}\left. -a \sqrt{2}(l(l+1)+2) \right)\h{te} = 0\,. 
\eea
It is possible to solve for $a$, $b$ and $\alpha$ in terms of $\beta$ and $\gamma$ in such a way that all the terms vanish separately. We get
\beq
a=\sqrt{\frac{2}{l(l+1)}} \beta - \sqrt{\frac{(l-1)(l+2)}{2l(l+1)}} \gamma\, ,
\label{afix}
\eeq
\beq
b=\sqrt{\frac{2}{l(l+1)}} \beta + \sqrt{\frac{2}{(l-1)l(l+1)(l+2)}} \gamma\, ,
\label{bfix}
\eeq
\beq
\alpha=\frac{2}{\sqrt{l(l+1)}} \beta + \frac{2}{\sqrt{(l-1)l(l+1)(l+2)}} \gamma\, .
\label{alphafix}
\eeq
Eq. (\ref{alphafix}) then selects a particular gauge choice for achieving this simplification.

Replacing (\ref{GravSreplace}), (\ref{afix}), (\ref{bfix}) y (\ref{alphafix}) in (\ref{GravSmodII}) and working with $\h{0r}$ as Lagrange multiplier allow us to obtain the following simple Lagrangian
\beq 
\lag^{II}_{l}=\frac{\gamma^2}{2}\left[\dot{h}^{te}_l\dot{h}^{te}_l - \drh{te}\drh{te} - l(l+1)\h{te}\h{te} \right] \, .
\label{GravSmod2lag}
\eeq
The corresponding Hamiltonian is
\beq
\ham^{II}_{l}= \pi^{te}_{l} \dot{h}^{te}_l - \lag^{II}_{l} = \frac{1}{2} \left[ \frac{\pi^{te}_{l}\pi^{te}_{l}}{\gamma^2r^2} + \gamma^2 r^2 \drh{te}\drh{te} + \gamma^2 l(l+1) \h{te}\h{te} \right]\,,
\eeq
with the canonical commutation relations
\beq
\left[\pi^{te}_{l}(t,r),h^{te}_{l}(t,r')\right]= i \delta(r-r') \, .
\eeq

Finally, by making the identifications
\beq
\phi^{II}_{l}=\gamma r \h{te} \quad , \quad P^{II}_{l}=\frac{\pi_{l}^{te}}{r\gamma}\,,
\eeq
the Hamiltonian of the scalar field modes is recovered in the form
\beq
\ham^{II}_{l}= \frac{1}{2} \left[ P^{II}_{l}P^{II}_{l} + \partial_r \phi^{II}_{l}\partial_r \phi^{II}_{l} + \frac{l(l+1)}{r^2} \phi^{II}_{l}\phi^{II}_{l} \right]\,,
\label{GravSSII}
\eeq
associated with the commutation relations
\beq
\left[P^{II}_{l}(t,r),\phi^{II}_{l}(t,r')\right]= i \delta(r-r')\, .
\label{GravScomII}
\eeq

\subsection{Analysis of the mode \texorpdfstring{$l=0$}{Lg}}
For the case $l=0$ the tensor spherical harmonics of spin $J=1$ y $J=2$ do not exist and the Lagrangian (\ref{GravSllagtot}) reduces to 
\bea
\lag_{l=0}= \lag_{l=0}^I + \lag_{l=0}^{II}= -\frac{r^2}{2} \alpha^2\dot{h}^{te}_0\dot{h}^{te}_0-\sqrt{2}r^2\alpha \dot{h}^{0l}_0\dot{h}^{te}_0+ \frac{r^2}{2}\alpha^2 \drho{te}\drho{te}  \nonumber \\
+\sqrt{2}\alpha r \ho{te}\drho{0l} +\ho{0l}\ho{0l}+2\sqrt{2}\ho{0r}\left[\sqrt{2}r\dot{h}^{0l}_0 -\alpha r^2 \partial_r \dot{h}^{te}_0 -  \alpha r \dot{h}^{te}_0\right] \label{GravSlag0} \\
+\sqrt{2}\ho{00} \left[-\sqrt{2}r\drh0{0l} -\sqrt{2}\ho{0l} +\alpha r^2 \partial_r\drho{te}+3\alpha r \drho{te}+ \alpha \ho{te} \right] \, . \nonumber
\eea
The equation of motion of $\ho{00}$ produces the constraint
\beq
-\sqrt{2}r\drho{0l} -\sqrt{2}\ho{0l} +\alpha r^2 \partial_r\drho{te}+3\alpha r \drho{te}+ \alpha \ho{te} =0 \, .
\label{GravSvinc0}
\eeq

By proposing the equivalent of (\ref{GravSreplace}) and replacing (\ref{GravSvinc0}) we get that the constants $a$ y $b$ must be $a=b=\alpha/\sqrt{2}$ without the need of fixing $\alpha$, in other words
\beq
\ho{0l}= \frac{\alpha}{\sqrt{2}}\left(\ho{te} +r \drho{te}\right) \quad \forall\alpha \, .
\label{GravSreplace0}
\eeq
On the other hand,  taking  $\ho{0r}$ as Lagrange multiplier gives
\beq
\sqrt{2}r\dot{h}^{0l}_0 -\alpha r^2 \partial_r \dot{h}^{te}_0 -  \alpha r \dot{h}^{te}_0=0\,.
\label{GravSvinc02}
\eeq
The equations (\ref{GravSreplace0}) and (\ref{GravSvinc02}) 
are clearly consistent with each other. Replacing both of them in (\ref{GravSlag0}) yields $\lag_{l=0}=0$, allowing us to conclude that the $l=0$ mode makes no contribution to the EE for any choice of gauge. 
\subsection{Analysis of the mode \texorpdfstring{$l=1$}{Lg}}
For the case $l=1$ the tensor spherical harmonics of spin $J=0$ y $J=1$ are well defined but the ones corresponding to $J=2$ do not exist. Hence the Lagrangian for the mode $I$  now writes  
\beq
\lag_{l=1}^I = \frac{r^2}{2} \dot{h}^{1m}_1\dot{h}^{1m}_1  +r^2\drhu{0m}\drhu{0m}+ 2\hu{0m}\hu{0m} +\sqrt{2}\dot{h}^{1m}_1\left( r\hu{0m} - r^2\drhu{0m}\right)\,. 
\label{GravSmodI1}
\eeq
In an analogous way to the case $l\geq 2$, we obtain $\pi_{1}^{1m} = r^2 \dot{h}^{1m}_1 + \sqrt{2}\left( r\hu{0m} - r^2\drhu{0m}\right) $ and the Hamiltonian can be expressed as
\beq
\ham^I_{l=1} = \frac{\pi_{1}^{1m}\pi_{1}^{1m}}{2r^2}-\sqrt{2}\hu{0m}\left(\partial_r \pi_{1}^{1m} + \frac{\pi_{1}^{1m}}{r} \right) \, .
\label{GravShamI0}
\eeq
Working with $\hu{0m}$ as a multiplier gives
\beq
\ham^I_{l=1} = \frac{\pi_{1}^{1m}\pi_{1}^{1m}}{2r^2}\,, \,\,\,\pi_{1}^{1m}= r \partial_r \pi_{1}^{1m} \, ,
\label{GravShamI02}
\eeq
which implies that the mode $I$ will not contribute to the EE for $l=1$. 

Moreover, the Lagrangian of mode $II$ can be given for $l=1$ from (\ref{GravSmodII}) as
\bea
\lag_{l=1}^{II} &=& \frac{r^2}{2} \left(\beta^2-\alpha^2\right)\dot{h}^{te}_1\dot{h}^{te}_1-\sqrt{2}r^2\alpha \dot{h}^{0l}_1\dot{h}^{te}_1+ \frac{r^2}{2}\alpha^2\drhu{te}\drhu{te} \nonumber \\
&+&\sqrt{2}\alpha r \hu{te}\drhu{0l} +\hu{0l}\hu{0l}+\left(\beta^2-\frac{\alpha\beta}{\sqrt{2}} \right) \hu{te}\hu{te}+\left( \sqrt{2}\alpha - 2\beta  \right)\hu{0l}\hu{te} \nonumber \\
&+&2\hu{0r}\hu{0r} +2\hu{0r}\left[ 2r\dot{h}^{0l}_1 -\sqrt{2}\alpha r^2 \partial_r \dot{h}^{te}_1 - \left(\sqrt{2} \alpha +\beta\right)r \dot{h}^{te}_1\right] \nonumber \\
&+&\hu{00} \left[-2r\drhu{0l} -4\hu{0l} +\sqrt{2}\alpha r^2 \partial_r\drhu{te}+\left(3\sqrt{2}\alpha+2\beta \right) r \drhu{te}+4\beta \hu{te} \right] \, . 
\label{GravSmodII1}
\eea
 so, $\hu{00}$ yields the constraint
\beq
-2r\drhu{0l} -4\hu{0l} +\sqrt{2}\alpha r^2 \partial_r\drhu{te}+\left(3\sqrt{2}\alpha+2\beta \right) r \drhu{te}+4\beta \hu{te} =0\,.
\label{GravSvincII1}
\eeq
In this calculation, we also propose the locality relation (\ref{GravSreplace}) and by replacing it in (\ref{GravSvincII1}) we obtain that for every choice of gauge it is valid that
\beq
\hu{0l}= \frac{\alpha}{\sqrt{2}} \hu{te} + r \beta \drhu{te}\quad \forall \, \alpha,\, \beta \, .
\label{GravSreplaceII1}
\eeq
Finally, using (\ref{GravSreplaceII1}) in (\ref{GravSmodII1}) produces $\lag_{l=1}^{II}=0 $. Thus, there is no contribution of the mode $II$ for $l=1$.

\subsection{Analysis of the gauge fixing}
We have already restricted the gauge choice with the relation (\ref{alphafix}) that allow us to write the dynamics of the two modes in the same fashion as the one of the scalar modes. Now we analyze if the field $h_{\mu\nu}$ or, more conveniently,  the resulting degrees of freedom associated with each mode $\h{1m}$ and $\h{te}$ can be written in terms of gauge invariant operators inside the sphere. For this purpose, we appeal to the expression (\ref{RDD}) of the gauge invariant curvature  tensor.

Using a computer based algebraic manipulation we obtain that the mode $I$ field given by  $\h{1m}$ can be rewritten in terms of the "electric-radial-electric-magnetic" contraction of the Riemann tensor 
\beq
R_{erem}^{lm}=e^\mu r^\nu e^\rho m^\sigma R_{\mu\nu\rho\sigma}^{lm} = F_{lm}(\theta,\varphi) \frac{\h{1m}(t,r)}{r^2} \,,
\label{erem}
\eeq
where $F_{lm}(\theta,\varphi) $ is a function of the angles $\theta$ and $\varphi$ for each $l$ and $m$. Specifically, for $m=0$ it is valid that
\beq
F_{l0}(\theta)=\frac{\pi^{\frac{5}{2}}l\sqrt{\Gamma^3(l+1)\Gamma(l)}}{16\Gamma^2(l+2)}\pl{0}[\pl{1}]^3\left(4\pl{2}\cot{\theta}+\pl{3}\right)\,,
\eeq
where $\pl{m}$ are the associated Legendre polynomials. The important point in this expression is that the relation between $\h{1m}(t,r)$ and the curvature does not involve radial derivatives. That would make the algebra generated by this field non local with respect to the one of gauge invariant operators in the sphere.

For the mode $II$, under the partial gauge choice (\ref{alphafix}), we can  further set  $\alpha=0$ or equivalently 
\beq
\gamma=-\frac{\beta}{\sqrt{(l-1)(l+2)}} \, ,\label{gabe}
\eeq
to obtain locality with respect to the curvature tensor. 
With this choice (\ref{GravSreplace}) reduces to an algebraic relation (without any derivatives) between the fields $\h{0l}$ y $\h{te}$ given by
\beq
\h{0l}=\sqrt{\frac{(l-1)(l+2)}{2}}\beta\h{te} \, .
\eeq
From this relation it follows that the remaining field $\h{te}$ can be computed from the "electric-magnetic-electric-magnetic" contraction of the Riemann tensor in a local way in $t,r$ as
\beq
R_{emem}^{lm}=e^\mu m^\nu e^\rho m^\sigma R_{\mu\nu\rho\sigma}^{lm} = G_{lm}(\theta,\varphi) \frac{h_{te}(t,r)}{r^2}\,,
\label{emem}
\eeq
where $G_{lm}(\theta,\varphi) $ is another function of the angles $\theta$, $\varphi$ for each $l$ and $m$. For $m=0$ it writes
\bea
G_{l0}(\theta)&=&\frac{\pi^{\frac{5}{2}}\beta\sqrt{l(l+2)\Gamma(l)}}{16(l+1)^2\sqrt{\Gamma(l+3)}}[\pl{1}]^4\left(4l(l+1)\pl{0}\right. \nonumber \\
&&+\left. 2(l(l+1)+2)\pl{1}\cot{\theta}+(l(l+1)+2)\pl{2}\right)\,.
\eea

Therefore, we conclude that, by taking $\alpha=0$, and eq. (\ref{gabe}) for the gauge fixing, the gauge fixed field $h_{\mu\nu}$ inside the sphere generates the same algebra as the gauge invariant operators. This algebra is equivalent to the one of the modes of two scalar fields except for the $l=0,1$ modes which are absent for the helicity $2$ theory. 

\subsection{Entanglement entropy and logarithmic coefficient}
To sum up, the EE associated with linearized gravitons in a sphere of radius $R$ is equivalent to the one corresponding to two scalar fields without contributions of the $l=0$ and $l=1$ angular momentum modes (or a Maxwell field without the $l=1$ modes). 

As we recall in section \ref{maxwellsphere}, the entanglement entropy of a scalar in a sphere has a universal logarithmic term $-1/90 \log(R/\epsilon)$ and the mode $l=0$ for the scalar corresponds to a massless $d=2$ scalar field in the $r>0$ half-line with entropy given by $1/6 \log(R/\epsilon)$. To obtain the universal logarithmic term for gravitons we just need the logarithmic contribution of the $l=1$ mode for the scalar. 

This mode is a $d=2$ field in the half-line $r>0$ with Hamiltonian
\be 
\ham= \frac{1}{2} \left[ P^2 + (\partial_r \phi)^2 + \frac{2}{r^2} \phi^2  \right]\,.
\ee
This model is scale invariant, but in contrast with the $d=2$ scalar field it contains a potential term $2/r^2 \phi^2$. We have to compute the entanglement entropy in an interval $r\in (0,R)$. The ultraviolet divergent piece of the EE comes from entanglement in high energy fluctuations around the boundary $r=R$. For these high energy fluctuations the effect of the potential can be neglected and then we must have a divergent piece that is the same as for the usual scalar field $S\sim -1/6 \,\log(\epsilon)$. As the model does not contain any dimensionfull scales by dimensional reasons we obtain
\be
S=  \frac{1}{6} \,\log(R/\epsilon)+ \textrm{cons}\,.
\ee
We have check this numerically in the lattice to an excellent (five digits) precision.

Hence, as for the $l=0$ mode, we get a $1/6$ coefficient for the logarithmic term of the $l=1$ modes. Consequently, summing up, we get a logarithmic coefficient for the graviton in the sphere given by twice the coefficient of the scalar subtracting two times the $l=0$ mode and $2 (2 l+1)=6$ times the $l=1$ mode, obtaining 
\be
2\times \left(-\frac{1}{90}-\frac{1}{6}-3 \times \frac{1}{6} \right)=-\frac{61}{45}\,.  
\ee
As it seems to be the rule, the value of the logarithmic coefficient increases with spin, being higher for helicity $2$ than for Maxwell and scalar fields. The entropy on the sphere then writes
\be
S=c \frac{A}{\epsilon^2}-\frac{61}{45}\, \log(R/\epsilon)\,.
\ee


\section{Discussion}
\label{dis}

We have computed the EE for free gravitons in flat space for a region between parallel planes and for the sphere. For the wall we find a universal coefficient that coincides with the one of two scalar fields. For the sphere the logarithmic term is given by $-61/45$, that is equivalent to two scalar fields where the $l=0$ and $l=1$ modes are missing. These results refer to clear physical quantities. First, our real time approach allow us to clarify that these are entropies of gauge invariant operator algebras of the theory inside the regions. Second, the meaning of these universal terms for the continuum model follows from the fact that they coincide with the ones obtained using the mutual information. We can write a regularized entropy as \cite{Casini:2015woa},
\be
S_\epsilon(A)\equiv \frac{1}{2}\, I_\epsilon(A_+,A_-)\,.    
\ee
In this formula one computes the mutual information between two regions $A_+$ and $A_-$ covering most of the inside and outside parts of the boundary of $A$ respectively, but symmetrically separated from the boundary by a distance $\epsilon/2$. This can be thought a  form of point splitting regularization of the entropy. The mutual information for disjoint regions is completely unambiguos in QFT and thus is $S_\epsilon(A)$. In particular, mutual information is unaffected by details of the algebra definition such as center terms (or edge modes). In the present case our results for the entropy are indeed equivalent to  $S_\epsilon(A)$. This is the case of the full scalar field EE \cite{Casini2016} and this identification also holds for the $l=0,1$ modes. These later one dimensional fields have mutual information that diverge as $-1/3 \log(\epsilon)$ as the boundaries of $A_+$ and $A_-$ approach each other. This holds for the free scalar and this UV result cannot change due to the potential or the boundary condition at the origin.\footnote{There is however a subleading $-1/2\log(\log(R/\epsilon))$ term in the mutual information for the $l=0$ mode that is not present in the entropy (with the usual lattice regularization) \cite{Casini2016}. This cames from superselection sectors for the $d=2$ scalar \cite{ss1,esecal}.}      

There are other results in the literature concerning the logarithmic coefficient due to gravitons, specially in black hole backgrounds (see for example \cite{Fursaev:1996uz,Solodukhin:2011gn,Sen:2012dw,Solodukhin:2015hma}; see also \cite{vassi} and references therein for gravitons in de Sitter space). There is the general expectation that the logarithmic coefficient for the sphere should be proportional to the $A$ anomaly\footnote{For black hole backgrounds another contribution is expected  proportional to the $c$ anomaly coefficient.} \cite{scalar,scalar2}. The free graviton does not have a symmetric gauge invariant stress tensor due to the Weinberg Witten theorem \cite{ww}, and then the definition of the $A$ anomaly is uncertain.\footnote{We thank a communication by Sergey Solodukhin regarding anomalies for the graviton.}  For a Maxwell field there is a  mismatch of the logarithmic term in the entanglement entropy and the $A$ anomaly which is solved by coupling the theory to (heavy) charges. In the present case a clarification of what is the right coefficient for interacting gravity seems to be further away since any interactions would take us away from the QFT setting, and thus rising the problems of operator algebra localization. Eternal black holes seem to be a more natural setup in gravity than the sphere since they are related to a partition of the asymptotic space in two. In this same sense, there are also indications that in full quantum gravity a boundary  separating localized degrees of freedom should be an extremal surfaces \cite{Camps:2018wjf,camps2}. This is of course the case of the entanglement wedge in holographic EE but not the sphere is Minkowski space.  

A natural conjecture that suggest itself from our results for the Maxwell field and the graviton is that on the sphere the EE of higher helicity $h>2$ fields should be equivalent to the one of two scalar fields where the $l=0,\cdots, h-1$ modes are subtracted. By the same reasons discussed in the previous section these modes have a EE given by  
\be
S=  \frac{1}{6} \,\log(R/\epsilon)+ \textrm{f(l)}\,,
\ee
where $f(l)$ is a function of the angular momentum. Hence, we would have a logarithmic coefficient\footnote{After this paper appeared in the arXiv database Dowker noted this same result would follow from thermodynamics in de Sitter space \cite{dodo}. He also obtains the result for fermion fields of different helicity.} 
\be
-2 \left(\frac{1}{90}+ \frac{1}{6} \sum_{l=0}^{h-1} (2 l+1)\right)=-\frac{1+15 h^2}{45}\,.
\ee

Another interesting problem is how to fix the gauge for the graviton in order that  $h^{\mu\nu}$ inside a region of arbitrary shape is given in terms of the gauge invariant operators localized in the same region.   We hope to come back to these problems in the future.

\section*{Acknowledgments} 
We thank discussions with Pablo Bueno, Joan Camps and  Marina Huerta. 
This work was partially supported by CONICET, CNEA
and Universidad Nacional de Cuyo, Argentina. The work of H. C. is partially supported by an It From Qubit grant by the Simons foundation. 

\appendix

\section{Properties of vector spherical harmonics}
\label{apa}

In this appendix we list some useful properties of vector spherical harmonics, some of then may also be found in \cite{Casini2016,Compere2018,Thorne1980}. The vector spherical harmonics are defined by (\ref{Ylmr},\ref{Ylme},\ref{Ylmm}).
They satisfy the orthogonality  relations
\beq
\int  \av{s} \av{s'*} d\Omega = \delta_{ss'} \delta_{ll'} \delta_{mm'}\,,
\eeq
where $\av{s*}$ is the complex conjugate of $\av{s}$, that is also given by
\beq
\av{s*}=(-1)^m \overline{Y}_{l(-m)}^s\, .
\eeq
The vector spherical harmonics can be used to expand and arbitrary three component vector $\overline{V}$ as
\begin{equation}
\overline{V} = \sum_{l=0}^\infty \sum_{m=-l}^l \sum_{s=r,e,m} V^s_{lm}(r) \av{s}(\theta,\varphi)
\end{equation}
where the functions $V^s_{lm}(r)$ are fixed by the Fourier coefficient expression 
\beq
V^s_{lm}(r) =\int  \overline{V} \cdot \overline{Y}^{s^*}_{lm} d\Omega \, .
\eeq
The vector spherical harmonics posses the following directional properties 
\beq
\hat{r} \cdot \av{r} = \as , \quad  \hat{r} \cdot \av{e} = 0, \quad  \hat{r} \cdot \av{m} = 0 \, .
\eeq
and their divergences are given by
\bea
\nabla \cdot \av{r} &=& \frac{2}{r} \as \, ,\\
\nabla \cdot \av{e} &=& -\frac{\sqrt{l(l+1)}}{r} \as \, ,\\
\nabla \cdot \av{m} &=& 0 \, .
\eea
The curls can be written as 
\bea
\nabla \cross \overline{Y}^{r}_{lm} &=& - \frac{\sqrt{l(l+1)}}{r} \overline{Y}^{m}_{lm} \, ,\\
\nabla \cross \overline{Y}^{e}_{lm} &=& \frac{1}{r} \overline{Y}^{m}_{lm} \, ,\\
\nabla \cross \overline{Y}^{m}_{lm} &=& - \frac{\sqrt{l(l+1)}}{r} \overline{Y}^{r}_{lm} - \frac{1}{r} \overline{Y}^{e}_{lm}\, .
\eea
Finally, the Laplancians can be computed to be
\bea
\nabla^2 \av{r} &=& - \frac{l(l+1)+2}{r^2} \av{r} + \frac{2\sqrt{l(l+1)}}{r^2}\av{e}\, ,\\
\nabla^2 \av{e} &=& \frac{2\sqrt{l(l+1)}}{r^2}\av{r}- \frac{l(l+1)}{r^2} \av{e} \, ,\\
\nabla^2 \av{m} &=& - \frac{l(l+1)}{r^2} \av{m} \, .
\eea

\section{Properties of tensor spherical harmonics}
\label{apb}
In this appendix we list some properties of tensor spherical harmonics, some of then may also be found in \cite{Compere2018,Thorne1980}. They are given by eqs. (\ref{22}) or alternatively by the expressions (\ref{33}). Their most useful characteristic is that they can be used as a basis for the space of symmetric tensors fields at a fixed radius. 
The tensor spherical harmonics satisfy the orthogonality relation 
\beq
\int  \traza{T_{lm}^{JS}{T^*}_{l'm'}^{J'S"}} d\Omega = \delta_{JJ'}\delta_{ss'}\delta_{ll'}\delta_{mm'}\, ,
\eeq
where ${T^*}_{lm}^{JS}$ is the complex conjugate of ${T}_{lm}^{JS}$ given by
\beq
{T^*}_{lm}^{JS}=(-1)^mT_{l-m}^{JS}\, .
\eeq
The traces of the tensor spherical harmonics are
\bea
\traza{T_{lm}^{0l}}=Y_{lm}\, , \qquad \traza{T_{lm}^{0t}}=\sqrt{2}Y_{lm} \,,\nonumber \\
\traza{T_{lm}^{Js}}=0\, , \qquad Js=1e,1m,2e,2m \, .
\eea
They further satisfy
\bea
 \hat{r} \cdot T_{lm}^{0l} = \overline{Y}_{lm}^r\, , \qquad \hat{r} \cdot T_{lm}^{1e} = \frac{1}{\sqrt{2}}\overline{Y}_{lm}^e \, ,\qquad  \hat{r} \cdot T_{lm}^{2e} = 0\, ,  \nonumber \\
\hat{r} \cdot T_{lm}^{0t} = 0\, , \qquad \hat{r} \cdot T_{lm}^{1m} = \frac{1}{\sqrt{2}}\overline{Y}_{lm}^m \, , \qquad  \hat{r} \cdot T_{lm}^{2m} = 0 \, .
\eea
The divergences of tensor spherical harmonics can be written as
\bea
\nabla \cdot \at{0l} &=&  \frac{2}{r}\av{r}\, ,\\
\nabla \cdot \at{0t} &=&-\frac{\sqrt{2}}{r}\av{r} + \frac{1}{r} \sqrt{\frac{l(l+1)}{2}}\av{e}\, ,\\
\nabla \cdot \at{1e} &=& - \frac{1}{r} \sqrt{\frac{l(l+1)}{2}}\av{r} + \frac{1}{r}\frac{3}{\sqrt{2}} \av{e}\, ,\\
\nabla \cdot \at{1m} &=&  \frac{1}{r}\frac{3}{\sqrt{2}} \av{m}\, ,\\
\nabla \cdot \at{2e} &=& -\frac{1}{r} \sqrt{\frac{(l-1)(l+2)}{2}} \av{e}\,, \\
\nabla \cdot \at{2m} &=& -\frac{1}{r} \sqrt{\frac{(l-1)(l+2)}{2}} \av{m} \, ,
\eea
and the Laplacians are the following
\bea
\nabla^2\at{0l} &=& -\frac{l(l+1)+4}{r^2} \at{0l} + \frac{2\sqrt{2}}{r^2} \at{0t} + \frac{2\sqrt{2l(l+1)}}{r^2} \at{1e}\, ,\\
\nabla^2 \at{0t} &=&  \frac{2\sqrt{2}}{r^2} \at{0l} -\frac{l(l+1)+2}{r^2} \at{0t} - \frac{2\sqrt{l(l+1)}}{r^2} \at{1e}\, ,\\
\nabla^2 \at{1e} &=& \frac{2\sqrt{2l(l+1)}}{r^2} \at{0l} - \frac{2\sqrt{l(l+1)}}{r^2} \at{0t}\nonumber  \\
&&-\frac{l(l+1)+4}{r^2} \at{1e} + \frac{2\sqrt{(l-1)(l+2)}}{r^2} \at{2e}\, ,\\
\nabla^2\at{1m} &=& -\frac{l(l+1)+4}{r^2} \at{1m} + \frac{2\sqrt{(l-1)(l+2)}}{r^2} \at{2m}\, ,\\
\nabla^2\at{2e} &=& \frac{2\sqrt{(l-1)(l+2)}}{r^2} \at{1e} - \frac{(l-1)(l+2)}{r^2} \at{2e}\, ,\\
\nabla^2 \at{2m} &=&\frac{2\sqrt{(l-1)(l+2)}}{r^2} \at{1m} - \frac{(l-1)(l+2)}{r^2} \at{2m}\,.
\eea

\end{document}